\documentstyle[preprint,aps,prd,psfig,here]{revtex}
\tightenlines
\newcommand{\be}{\begin{equation}}
\newcommand{\ee}{\end{equation}}
\newcommand{\bea}{\begin{eqnarray}}
\newcommand{\eea}{\end{eqnarray}}
\newcommand{\bean}{\begin{eqnarray*}}
\newcommand{\eean}{\end{eqnarray*}}
\newcommand{\ba}{\begin{array}}
\newcommand{\ea}{\end{array}}

\begin{document}
\draft
\title{Optimal Spin Basis \\
       in Polarized Photon Linear Colliders}
\author{
Michihiro HORI, 
\thanks{E-mail:michi@jigen.phys.sci.hiroshima-u.ac.jp},
Yuichiro KIYO
\thanks{E-mail:kiyo@theo.phys.sci.hiroshima-u.ac.jp}
and Takashi NASUNO 
\thanks{E-mail:nasuno@theo.phys.sci.hiroshima-u.ac.jp}
}
\date{\today}
\address{Dept. of Physics,Hiroshima University \\Higashi-Hiroshima 739,Japan }
\preprint{HUPD-9712, hep-ph/9712379}
\maketitle
\vspace{3cm}
%
%%%%%%%%%%%%%%%%%%%%%%%%%%%%%%%%%%%%%%%%%%%%%%%%%%%%%%
\begin{abstract}
%%%%%%%%%%%%%%%%%%%%%%%%%%%%%%%%%%%%%%%%%%%%%%%%%%%%%%
%
We analyze the spin correlations of the top quark pairs 
produced at Photon Linear Colliders. We employ the circular 
polarized photon beams and general spin basis for the 
top quark pair. We consider general spin bases to find a 
strong spin correlation between produced top quark and 
anti-top quark. We show the cross-section in these bases and 
discuss the characteristics of results.
\end{abstract}
%
%%%%%%%%%%%%%%%%%%%%%%%%%%%%%%%%%%%%%%%%%%%%%%%%%%%%%%
%                PACS number
\vspace{3cm}
%%%%%%%%%%%%%%%%%%%%%%%%%%%%%%%%%%%%%%%%%%%%%%%%%%%%%%
%
%%%%%%%%%%%%%%%%%%%%%%%%%%%%%%%%%%%%%%%%%%%%%%%%%%%%%%
\section{Introduction}
%%%%%%%%%%%%%%%%%%%%%%%%%%%%%%%%%%%%%%%%%%%%%%%%%%%%%%
%
We have been progressing in our understanding of nature step by step. 
LEP and Tevatron experiments provided us with the precision measurement 
of the top quark mass \cite{CDF,D0}, W boson mass, the bound for the 
Higgs boson mass and many other electro-weak parameters. 
With these numerous data, the standard electro-weak theory are enjoying 
the nature without any discrepancies at least in current experiments.

In 1994, the last fundamental quark of the ``Standard Model'', 
the top was discovered at the Fermi-Lab Tevatron.
The discovery of this very heavy fundamental particle is remarkable, 
which brought us some good chance to probe the electro-weak symmetry 
breaking mechanism.
The top quark sector is not yet well established and many authors are 
discussing related problems. For instance, T or CP violation in the 
top quark sector and the possibility of the presence of anomalous couplings 
have been discussed in many papers \cite{SEE}.
A very heavy top quark arouse our interests in trying to understand 
the nature deeply and we expect to investigate the top quark sector of
the Standard Model at the Next Linear Collider (NLC) \cite{NLC1,NLC2}
in the near future. 

The top quark physics at NLC is attractive and will give us some new 
information to understand the Standard Model and clue to physics beyond 
the Standard Model.

On the other hand, the Photon Linear Colliders \cite{OHGA} may be the best 
alternatives to the electron positron colliders. 
Physics opportunities in the Photon Linear Colliders are as rich as 
those in the $e^{+} e^{-}$ colliders and furthermore we have an unique 
opportunity in the case of the Photon Linear Colliders, namely, we 
can control the initial photon polarization by the inverse Compton scattering 
of the polarized laser by the electron/positron beams at NLC. 
Handling the polarized photon beams, we have the total angular momentum 
$J = 0, 2, \cdots$ states in the s-channel.
Using these polarized high energy photon beams, we will have attractive 
channels for Higgs particle production \cite{OHGA}, top quark pair production  
\cite{GABB} and other interesting processes.

It was discussed in the papers \cite{JEZA,KUHN,GREG} that the spin of the top 
quark can be determined from the angular distribution of the electro-weak 
decay products of the top quark. 
So, it is interesting to investigate the spin correlations of the top quark
pairs at the Photon Linear Colliders.
We discuss in this paper the top quark pair production at the Photon Linear 
Colliders with the circular polarized photon beams and the general
spin basis for the top and anti-top quarks.

This paper is organized as follows.
In Sec.II we explain our notation and convention to calculate the 
cross-section. We show the cross-section of the top quark pair production
from the initially polarized two photon beams. 
In Sec.III we show that there is an useful spin basis to investigate the
spin correlations of the top quark pairs at Photon Linear Colliders, 
which was first discussed by Mahlon and Parke \cite{GREG} at hadron colliders 
and was extended by Parke and Shadmi \cite{PARK} to the $e^{+} e^{-}$ annihilation
process.

Because of the large mass of the top quark, it decays through the electro-weak 
process before the QCD hadronization effects come in \cite{BIGI}. 
So the decay products are the messengers of the top spin.
In that sense, decay products are sometimes referred to as the spin analyzer 
of the top quark. We briefly review this point in Sec.IV. 
Sec.V is devoted to the summary.
Some similarities between the gluon-gluon fusion and the photon-photon
fusion into the top quark pair are also mentioned.
Some formulae needed in the calculation with the spinor helicity basis 
are collected in Appendix A. 

%
%%%%%%%%%%%%%%%%%%%%%%%%%%%%%%%%%%%%%%%%%%%%%%%%%%%%%%
\section{Production of Polarized Top }
%%%%%%%%%%%%%%%%%%%%%%%%%%%%%%%%%%%%%%%%%%%%%%%%%%%%%%

We present the cross-section for the polarized top-quark 
pair production from the circular polarized two photons 
($\gamma_{R,L},\gamma_{R,L}\rightarrow t_{\uparrow,\downarrow}
\bar{t}_{\uparrow,\downarrow}$) in the center-of-mass (CM) frame 
at the tree level in the perturbation theory [Fig.1(a)]. 
The suffix $\uparrow/\downarrow$ denotes the spin up/down for 
the top and anti-top quarks and the state $\gamma_{R}/\gamma_{L}$ 
refers to the right-handed/left-handed photon. 

At the NLC, we have high energy polarized photon beams which will be 
produced by the inverse Compton scattering. Adjusting the laser beam 
polarization, we can get highly polarized photon beams \cite{NLC1}. 
The linear polarized beams will be also available but in this paper 
we discuss only the case of circular polarized beams to get clear  
information on the top quark spin.

Since we don't discuss the T-violating interaction in the photon and
top quark coupling, we can choose the top and anti-top spins to lie
on the production plane. Furthermore we introduce only one parameter
$\xi$ to define the spins of the top and anti-top quarks.
This makes the expression of the cross-section simple and we can relate
some amplitudes each other. 
The definition of $\xi$ is as follows \cite{GREG,PARK}:
We decompose the top spin along the direction $\vec{s}_{t}$ in the 
rest frame of the top quark [Fig.1(b)], where it has a relative 
angle $\xi$ clock-wisely to the anti-top quark momentum. 
Thus the state $t_{\uparrow}$/$t_{\downarrow}$ refers to the top 
with spin in the direction $+\vec{s}_{t}/-\vec{s}_{t}$ respectively. 
Similarly the anti-top spin is defined along the direction $\vec{s}_{\bar{t}}$ 
in the anti-top quark rest frame [Fig.1(c)], having the same angle $\xi$ 
clock-wisely from the top quark momentum.
The state $\bar{t}_{\uparrow}$/$\bar{t}_{\downarrow}$ refers to the anti-top
with spin in the direction $+\vec{s}_{\bar{t}}/-\vec{s}_{\bar{t}}$ respectively. 
Thus all momenta and spin vectors $\vec{s}_{t},\vec{s}_{\bar{t}}$ in Fig.1 
are lying on the production plane.

In our calculations, we employ the spinor helicity method described in
Refs. \cite{KLEI,MANG}. The definitions of the massive spin state, photon 
helicity state are summarized in the Appendix A.

The cross-sections in the CM frame are the functions of the scattering 
angle $\theta^{\ast}$, the top (anti-top) quark speed $\beta$ and 
the angle $\xi$ which defines the spin axis [Fig.1].
When the initial two photons have the total angular momentum $J=2$,
we get, 
\bea
\frac{d\sigma}{d\cos\theta^{\ast}}
\left(
\gamma_{R}~ \gamma_{L} 
\rightarrow t_{\uparrow} \bar{t}_{\uparrow} ~\mbox{and}~
t_{\downarrow}\bar{t}_{\downarrow}
\right) 
&=&
y\left( \beta ,\theta^{\ast} \right)
\times 
\beta^{2} \sin^{2}\theta^{\ast}
\nonumber \\
&\times &
\left(
\sqrt{1-\beta^{2}}\sin\theta^{\ast}\cos\xi 
- 
\cos\theta^{\ast} \sin\xi 
\right)^{2},
\label{eqn:RLUU}
\eea
\bea
\frac{d \sigma}{d\cos\theta^{\ast}}
\left(
\gamma_{R}~ \gamma_{L} 
\rightarrow t_{\uparrow} \bar{t}_{\downarrow} ~\mbox{or}~
t_{\downarrow}\bar{t}_{\uparrow}
\right) 
&=&
y\left( \beta ,\theta^{\ast} \right)
\times
\beta^{2}\sin^{2}\theta^{\ast}
\\ 
\nonumber 
&\times &
\left(
\sqrt{1-\beta^{2}}\sin\theta^{\ast}\sin\xi
+
\cos\theta^{\ast}\cos\xi\mp 1
\right)^{2}.
\label{eqn:RLUD}
\eea
The function $y(\beta,\theta^{\ast})$ is a common factor and 
has no $\xi$ dependence,
\bea
&&
y\left( \beta ,\theta^{\ast} \right)
\equiv
\frac{\beta}{32 \pi s} 
\times
\frac{
4 N_{C}\left(4\pi\alpha_{QED}\right)^{2} 
Q_{t}^{4}
}{
\left(1- \beta^{2} \cos^{2}\theta^{\ast}
\right)^{2}
}.
\label{eqn:Y-FUNC}
\eea
Here $\sqrt{s}$ is the CM energy, $\beta =\sqrt{1-4m_{t}^2/s}$, 
$\alpha_{QED}$ is the QED fine structure constant, $Q_{t}=2/3$ and
$N_{C}$ is the number of color.

For the initial state which has the total angular momentum $J=0$, we 
obtain,
\bea
\frac{d\sigma}{d\cos\theta^{\ast}}
\left(
\gamma_{R}~ \gamma_{R} 
\rightarrow t_{\uparrow} \bar{t}_{\uparrow} ~or~
t_{\downarrow}\bar{t}_{\downarrow}
\right) 
&=&
y\left( \beta ,\theta^{\ast} \right)
\times
\left(1-\beta^{2}\right) 
\left(
1\mp \beta \cos\xi
\right)^{2},
\label{eqn:RRUU}
\\
\frac{d\sigma}{d\cos\theta^{\ast}}
\left(
\gamma_{R}~ \gamma_{R} 
\rightarrow t_{\uparrow} \bar{t}_{\downarrow} ~and~
t_{\downarrow}\bar{t}_{\uparrow}
\right) 
&=&
y\left( \beta ,\theta^{\ast} \right)
\times
\left(1-\beta^{2}\right) 
\beta^{2}\sin^{2}\xi.
\label{eqn:RRUD}
\eea
The other cross-sections, 
$d \sigma(\gamma_{L}\gamma_{R}\rightarrow t \bar{t})/d\cos\theta^{\ast}$ or 
$d \sigma(\gamma_{L}\gamma_{L}\rightarrow t \bar{t})/d\cos\theta^{\ast}$ 
can be obtained from Eqs.(\ref{eqn:RLUU})-(\ref{eqn:RRUD})
by interchanging $L$ and $R$ as well as $\uparrow$ and $\downarrow$.

Before discussing the differential cross-sections, let us make some
comments on the general behavior of the total cross-sections in each
channel $J=0$ and $J=2$.
We show our results in Fig.2, summed over the top and anti-top 
spins in each channel of $J=0$ and $J=2$. We use the value $m_{t}=175$ 
GeV. The realistic energy of the Photon Linear Collider to produce the 
top pairs might be $\sqrt{s}=400 \sim 500$ GeV, so we have shaded the
corresponding region in figures. The behavior of the total
cross-sections is controlled by the factor in the expressions of 
the cross-sections for each channel, $\beta^{2}$ for the $J=2$ channel 
and $(1-\beta^{2})$ for the $J=0$ channel. 
In the $J=2$ channel, top pair production is suppressed when $\beta$ 
goes to $0$, namely in the Non-Relativistic (NR) Limit. 
The NR limit corresponds to the situation in 
which the top and anti-top quarks are produced at rest. 
So the total angular momentum $J=2$ state can not be realized 
and the cross-section is suppressed according to the angular 
momentum conservation. 
The factor $(1-\beta^2)$ in $J=0$ channel suppresses the 
cross-section in the Ultra-Relativistic (UR) limit\footnote{
The higher order corrections might change these situations 
slightly, especially the behavior of the cross-section for 
$J=0$ \cite{JIKI}.}.
Actually one can see these behavior in Fig.2.
%
%%%%%%%%%%%%%%%%%%%%%%%%%%%%%%%%%%%%%%%%%%%%%%%%%%%%%%%%%%%%%%%%%%%%%%
\section{Optimal Spin Basis}
%%%%%%%%%%%%%%%%%%%%%%%%%%%%%%%%%%%%%%%%%%%%%%%%%%%%%%%%%%%%%%%%%%%%%%
In this section, we show that there is an optimal spin basis 
which maximizes the spin correlations between the produced top and
anti-top quarks. These spin correlations may be useful when one 
wants to distinguish the Standard Model from some other models.

At first we discuss optimal spin basis for $J=0$ channel.
The most familiar spin basis might be the helicity basis which is 
defined by, 
\bea
\cos \xi & = & \pm 1.
\label{eqn:HELI}
\eea
In this basis, the cross-section for $\gamma_{R} \gamma_{R} \rightarrow t
\bar{t}$ becomes very simple and the spin correlation of the top quark pair 
is very strong. 
So in this channel, we employ the familiar helicity basis.
The cross-sections for the like-spin configurations, 
$t_{R}\bar{t}_{R}$ and $t_{L}\bar{t}_{L}$, are given by    
\bea
\frac{d\sigma}{d\cos\theta^{\ast}}
\left(
\gamma_{R}~ \gamma_{R} 
\rightarrow t_{R} \bar{t}_{R} ~or~
t_{L}\bar{t}_{L}
\right) 
&=&
y\left( \beta ,\theta^{\ast} \right)
\times
\left(1-\beta^{2}
\right) 
\left( 1 \pm \beta 
\right)^{2},
\label{eqb:RR-LIKE-H}
\eea
and for the unlike spin configurations in the $J=0$ 
channel are identically zero in this basis,
\bea
\frac{d\sigma}{d\cos\theta^{\ast}}
\left(
\gamma_{R}~ \gamma_{R} 
\rightarrow t_{R} \bar{t}_{L} ~and~
t_{L}\bar{t}_{R}
\right) 
&=&
0,
\label{eqn:RR-LIKE-H}
\eea
where and in what follows $t_{R/L}(\bar{t}_{R/L})$ refers to the top 
(anti-top) with spin up/down in ``helicity basis with $\cos\xi=-1$''.  

In Fig.3 we show the fraction $\sigma(t_{s}\bar{t}_{s'})/\sigma_{J=0}$ 
where $ \sigma_{J=0} = \sum_{s,s'}\sigma(t_{s}\bar{t}_{s'})_{J=0}$.
We plot the differential cross-section in Fig.4 at $\sqrt{s}=400$
GeV. The $t_{L} \bar{t}_{L}$ cross-section is strongly suppressed by the
factor $(1-\beta)^2$ compared to the $t_{R} \bar{t}_{R}$ as $\beta$
increases. Roughly speaking, the ratio is 
$d \sigma \left( t_{R} \bar{t}_{R} \right): d \sigma \left( t_{L}
\bar{t}_{L} \right) = 8:1$ 
at $\sqrt{s}=400$ GeV ($\beta \simeq 0.48 $).

In the $J=0$ channel the forward and backward scattering dominate the 
cross-section because the factor $y(\beta,\theta^{*})$ becomes large 
at $\theta^{*}=0,\pi$ when $\beta$ goes to 1.
While for $J=2$ channel forward and backward scattering are zero because 
the factor $\sin^{2}\theta^{*}$ cancels the singular behavior of the 
$y(\beta,\theta^{*})$ at $\theta^{*}=0,~\pi$. This characteristic
behavior
can be seen in the differential cross-sections.

Now we come to the discussion of the spin basis for the process
$\gamma_{R}\gamma_{L} \rightarrow t \bar{t}$ in which the initial
angular momentum is $ J =2 $.
The cross-sections in this channel with the helicity basis are 
\bea
\frac{d\sigma}{d\cos\theta^{\ast}}
\left(
\gamma_{R}~ \gamma_{L} 
\rightarrow t_{R} \bar{t}_{R} ~\mbox{and}~
t_{L}\bar{t}_{L}
\right) 
&=&
y\left( \beta ,\theta^{\ast} \right)
\times
\beta^{2} \sin^{2}\theta^{\ast}
(1-\beta^2)\sin^{2}\theta^{\ast},
\label{eqn:RL-LIKE-H}
\\
\frac{d \sigma}{d\cos\theta^{\ast}}
\left(
\gamma_{R}~ \gamma_{L} 
\rightarrow t_{R} \bar{t}_{L} ~\mbox{or}~
t_{L}\bar{t}_{R}
\right) 
&=&
y\left( \beta ,\theta^{\ast} \right)
\times
\beta^{2}\sin^{2}\theta^{\ast}
\left(\cos\theta^{\ast} \pm 1\right)^{2}.
\label{eqn:RL-UNLIKE-H}
\eea
It is natural to choose the spin basis as the helicity 
basis in the UR limit. However the top quark is so heavy that 
we do not have to stick to the helicity basis. 
A different choice of the spin basis will be very useful which 
maximizes the spin correlations between the top and anti-top quarks. 
In fact, it has been known that there exists the ``off-diagonal
basis'' which makes the contribution from the like spin configuration 
to vanish for the $e^{+}e^{-} \rightarrow t\bar{t}$ process \cite{PARK}. 
Here we show that even in the process 
$\gamma_{R}\gamma_{L} \rightarrow t\bar{t}$, we can take the 
off-diagonal basis by defining the spin angle $\xi$ as follows,
\bea
\tan\xi
& = &
\sqrt{1-\beta^2} \tan\theta^{\ast}.
\label{eqn:OD-BASIS}
\eea

We get the following expressions in this basis, 
\bea
\frac{d \sigma}{d\cos\theta^{\ast}}
\left(
\gamma_{R}~ \gamma_{L} 
\rightarrow t_{U} \bar{t}_{U} ~\mbox{and}~
t_{D}\bar{t}_{D}
\right) 
&=&
0 ,
\label{eqn:RL-UNLIKE-OD} \\
\frac{d\sigma}{d\cos\theta^{\ast}}
\left(
\gamma_{R}~ \gamma_{L} 
\rightarrow t_{U} \bar{t}_{D} ~\mbox{or}~
t_{D}\bar{t}_{U}
\right) 
&=&
y\left( \beta ,\theta^{\ast} \right)
\times
\beta^{2} \sin^{2}\theta^{\ast}
\left(
1 \mp \sqrt{1-\beta^{2}\sin^{2}\theta^{\ast}}
\right)^{2}.
\label{eqn:RL-LIKE-OD}
\eea
The state $t_{U/D}(\bar{t}_{U/D})$ refers to the top 
(anti-top) with spin up/down in the ``off-diagonal basis''.
We use this notation to distinguish the off-diagonal basis 
from helicity basis. 

To see the difference between the helicity and off-diagonal 
bases, we plot the fraction normalized by 
$\sigma_{J=2} = \sum_{s,s'}\sigma(t_{s}\bar{t}_{s'})_{J=2}$,
as a function of the speed $\beta$ [Fig.5]. 
Note that following equations hold, 
$\sigma(t_{R}\bar{t}_{L})_{J=2}=\sigma(t_{L}\bar{t}_{R})_{J=2}$
and 
$\sigma(t_{R}\bar{t}_{R})_{J=2}=\sigma(t_{L}\bar{t}_{L})_{J=2}$.
So the sum of them are plotted in Fig.5.

In the helicity basis, all spin configurations contribute to the 
cross-section $\sigma_{J=2}$ in a broad energy region, 
while one configuration $t_{D}\bar{t}_{U}$ dominates the 
cross-section in the off-diagonal basis.
The cross-section for $t_{U}\bar{t}_{D}$ is small when 
$ \beta < 1$ because it is proportional to $\beta^{4}$. 
At very high energy, one particular spin configuration dominates the
cross-section in both bases. 
Please note that the spin angle $\xi$ depends on the scattering angle 
$\theta^{*}$ and speed $\beta$ and the off-diagonal basis reduces to 
the helicity basis in the UR limit.
This is natural because the helicity basis is relevant in the 
high energy process.

In Fig.6 we show the differential cross-sections for each spin 
configuration at $\sqrt{s}=400$ GeV. As we have already mentioned, 
the spin configuration down-up $t_{D}\bar{t}_{U}$ dominates the 
cross-section in the off-diagonal basis in contrast with the cross-section
in the helicity basis. 
Therefore we can uniquely determine the spin configuration of top 
and anti-top quark to be ``down-up (DU)'' in the off-diagonal basis.

%
%%%%%%%%%%%%%%%%%%%%%%%%%%%%%%%%%%%%%%%%%%%%%%%%%%%%%%%
\section{Decay Products and top quark spin}
%%%%%%%%%%%%%%%%%%%%%%%%%%%%%%%%%%%%%%%%%%%%%%%%%%%%%%%
%
We will discuss the measurement of the top quark spin in this section.
Our discussion is all based on the Standard Model.
Thanks to the heavy mass of the top quark, it will decay before the 
QCD interaction comes in.
Thus we can obtain the information of the top quark spin by measuring 
it's decay products without being suffered from the complicated
hadronization effects \cite{BIGI}. 

To utilize the hadronic decay products as a spin analyzer, 
we must take into account the efficiency to identify the 
d-type quark from u-type (or vice versa) in the process 
$ t \rightarrow b \bar{d} u $ \cite{GREG}. Therefore we consider the 
the semi-leptonic decay of the top (anti-top) quark 
($t \rightarrow b \bar{l}\nu ~,~ \bar{t}\rightarrow\bar{b}l\bar{\nu} $) 
to simplify our discussions.

To see how the information of the top quark spin are transmitted to the 
decay products in the processes $t \rightarrow b \bar{l} \nu$, remember
that the longitudinal $W_{L}$ dominates in the decay process $t \rightarrow b W$.  
(When the longitudinal polarization vector $\epsilon^{\mu}_{L}(k)$ 
becomes parallel to its momentum $k^{\mu}$, the contribution of 
$W_{L}$ has a large factor $m_{t}/m_{w}$, $m_{w}$ is the W boson mass.)

The direction of the $W_{L}$ boson momentum is that of the 
top quark spin due to the angular momentum conservation.   
In the subsequent decay, the leptons $\bar{l}$ emitted from $W_{L}$ 
are highly boosted. Hence the leptons $\bar{l}$ can go away into 
the direction of the top quark spin.
Similar analysis for the anti-top quark tells us that the favorable 
direction of the momentum $l$ is the opposite to the direction of 
the anti-top quark spin.

A quantitative analysis yields the following distribution for the
decay product from the top in the rest frame \cite{JEZA,KUHN,GREG}.
\bea
\frac{1}{\Gamma_{T}}
\frac{d\Gamma}{d \cos\theta_{i}}
& = &
\frac{1 + \alpha_{i} \cos\theta_{i}
     }{2},
\label{eqn:DECAY-DIS}
\eea
where $\theta_{i}$ denotes the angle between the top quark spin axis
$\vec{s}_{t}$ and the decay product ($i=b, \bar{l}, \nu, u, \bar{d}$) and 
$\alpha_{i}$ depends on which particle we choose as a spin analyzer\footnote{
        $\alpha_{\bar{l}}=1$ for the charged lepton from the top quark whose
        spin is up $t(\uparrow)$ and $\alpha_{\bar{l}}=-1$ from the top
        quark whose spin is down $t(\downarrow)$. 
        When we choose another particle as a spin analyzer, the efficiency
        decreases (e.g. $\alpha_{\nu} \sim -0.31$ for $\nu$) \cite{JEZA,KUHN,GREG}.
        }.
The most efficient spin analyzer is the charged lepton, for which 
$\alpha_{\bar{l}}=1$.
From Eq.(\ref{eqn:DECAY-DIS}), we can obtain the probability 
$P\left(\uparrow|~\mbox{in the cone}\right)$\footnote{
        The probability function $P(A|B)$ denotes
        the conditional probability of the event A.
        }
that the top quark has spin up when we pick up events which satisfy the
condition $ \cos\theta_{i}> y $,
\bea
P
\left(\uparrow|~\mbox{in the cone}
\right)  
& = &
\frac{
\left(1 + \left < s_{t} \right > \right) \left( 2+\alpha_{i}(1+y)
\right)
      }
{4+ 2 \left< s_{t} \right> \alpha_{i} (1+y)},
\label{eqn:PROB-UP}
\eea
where $\left< s_{t} \right>$ denotes the average value of the top quark spin
defined by 
\bea
\left<s_{t}\right>
= 
\sum_{s_{t}=-1,1} s_{t} \times P(s_{t}).
\eea
$P(s_{t})$ is the probability that the top quarks with spin $s_{t}$
are produced.

It is straightforward to extend this analysis to the top pair
production process. The double decay distribution for the decay
product ``$i$'' from the top quark and the decay product``$\bar{i}$'' 
from the anti-top quark is given by 
\bea
\frac{d^{2}\Gamma }{d(\cos\theta_{i}) d(\cos\theta_{\bar{i}}) }
\sim
\frac{1}{4}
\left \{ 
1 + \alpha_{i} \left< s_{t} \right> \cos\theta_{i}
+ \alpha_{\bar{i}} \left< s_{\bar{t}} \right> \cos\theta_{\bar{i}}
+ \alpha_{i}\alpha_{\bar{i}}
\left< s_{t}s_{\bar{t}} \right> \cos\theta_{i}\cos\theta_{\bar{i}} 
\right \}, 
\label{eqn:DOUB-DECAY-DIS}
\eea
where $\theta_{i}$ $(\theta_{\bar{i}})$ is the angle between 
the direction of the top (anti-top) quark spin and the momentum of
decay product $i$ ($\bar{i}$) in the top quark (anti-top quark) 
rest frame.
As can be seen in the above expression, it is obvious how the spin 
correlation between the top and anti-top quarks is related 
to the distribution of their decay products. 

When we collect events for which the charged lepton $i$ from
the top quark lies in the cone defined by $\cos\theta_{i}>y$, 
the effective $\alpha$ in the expression Eq.(\ref{eqn:DECAY-DIS}) 
for anti-top quark becomes,
\bea
\alpha^{eff}_{\bar{i}}
& = &
\frac{ 
\{ 
2 \left < s_{\bar{t}} \right > + \alpha_{i} \left < s_{t}s_{ \bar{t}}
\right > (1+y) 
\}
      }{ 2+\alpha_{i} \left <s_{t}\right > (1+y) }
\times \alpha_{\bar{i}},
\label{eqn:EFF-ALPHA}
\eea
which determine the $\bar{i}$ distribution under the condition 
mentioned above.
Then the distribution of the decay product $\bar{i}$ from the 
anti-top quark is given by
\bea
\frac{1}{\Gamma_{T}}
\frac{d \Gamma
      }{ d \cos\theta_{\bar{i} }}
& = &
\frac{
1 + \alpha^{eff}_{\bar{i} } 
    \cos \theta_{\bar{i} }
     }{2}.
\label{enq:DECAY-DIS-C}
\eea
By observing this distribution, we can obtain the averaged value of the
spin $ \left <s_{t}\right >$ and the spin correlation 
$\left<s_{t}s_{\bar{t}}\right>$ between the top and anti-top quarks.

When one takes the partial spin average of initial photons,
$\gamma_{R}\gamma_{L}+\gamma_{L}\gamma_{R}$ or 
$\gamma_{L}\gamma_{L}+\gamma_{R}\gamma_{R}$,
which gives $\left <s_{t}\right >=\left <s_{\bar{t}}\right >=0$ 
and Eq.(\ref{eqn:EFF-ALPHA}) reduce to 
\bea
\alpha^{eff}_{\bar{i}}
& = &
\frac{ 
\alpha_{i} \left < s_{t}s_{ \bar{t}} \right > (1+y) 
      }{ 2 }
\times \alpha_{\bar{i}}.
\label{eqn:EFF-ALPHA-SIMP}
\eea
In this case the $\alpha_{\bar{i}}^{eff}$ gives the spin 
correlations between top and anti-top quarks directly. 

%
%%%%%%%%%%%%%%%%%%%%%%%%%%%%%%%%%%%%%%%%%%%%%%%%%%%%%%%
\section{Summary}
%%%%%%%%%%%%%%%%%%%%%%%%%%%%%%%%%%%%%%%%%%%%%%%%%%%%%%%

The top quark pair production process at Photon Linear Colliders
is discussed at the tree level in the perturbation theory. 
We have shown the analytic cross-section.
We focused on the spin correlation between the top and the anti-top
quarks and the spin structure of the cross-section.

It is worth to mention the cross-section for gluon-gluon 
fusion into the top pair. 
It is interesting that the squared amplitudes are completely 
the same as the one for the process $\gamma \gamma \rightarrow
t \bar{t}$ up to the function which does not have $\xi$ dependence.  
This process will be important to study the high energy top quark 
pair production in the future high energy polarized hadron colliders.
We present the analytic form of the squared amplitude in the Appendix B.

In conclusion, in the $J=0$ channel, the helicity basis is a good basis.
The dominant component of the signal is $t_{R}\bar{t}_{R}$, which makes 
up more than $89\%$ of the total cross-section at $ \sqrt{s}= 400 $ GeV.

In the $J=2$ channel, we present the polarized cross-sections
in the helicity and off-diagonal bases.
In the helicity basis, the total cross-section is not dominated by one
particular spin configuration.
In the off-diagonal basis, in contrast, only one spin configuration is
appreciably different from zero for all values of the scattering angle.
So we summarize that the off-diagonal basis is the most useful spin basis 
to find the strong spin correlations between produced top and
anti-top quarks.

This spin correlation can be measured by analyzing the decay products of
the top quark and anti-top quark. To observe these spin correlations
at the Photon Linear Colliders is interesting and may be a good test 
for the top quark sector of the Standard Model.

%%%%%%%%%%%%%%%%%%%%%%%%%%%%%%%%%%%%%%%%%%%%%%%%%%%%%%%
\section{Acknowledgment}
%%%%%%%%%%%%%%%%%%%%%%%%%%%%%%%%%%%%%%%%%%%%%%%%%%%%%%%
The authors would like to thank S. Parke for useful 
suggestions and comments. We thank J. Kodaira for carefully
reading the manuscript and encouraging us in the course of 
this work.
We would like to thank T. Takahashi for useful comments and  
the information about Photon Linear Colliders.

\clearpage
%%%%%%%%%%%%%%%%%%%%%%%%%%%%%%%%%%%%%%%%%%%%%%%%%%%%%%%
\section{Appendix A}
%%%%%%%%%%%%%%%%%%%%%%%%%%%%%%%%%%%%%%%%%%%%%%%%%%%%%%%
In this Appendix, we summarize the fermion spin state and the photon
helicity state.
We follow the conventions and notations defined in Refs \cite{GREG,MANG}.
At first, we define the chiral projections as follows:
%---------------------------------------------------------
\bea
\omega _{+} & \equiv & 
        \frac{1}{2}( 1 + \gamma _{5} ),~
\omega _{-} ~ \equiv ~ 
        \frac{1}{2}( 1 - \gamma _{5} ).
\eea
%---------------------------------------------------------
We use the following notation and conventions for the massless 
helicity states.
\bea
\left |p,\pm \right > &=& u_{\pm}(p) ~ = ~ \omega _{\pm} u(p),
\\
\left < p,\pm \right | &=& \bar{u}_{\pm}(p).
\eea
%---------------------------------------------------------
A state of the massive particle with momentum $p$ ($p^{2}= m^{2}$) is expressed 
as a superposition of the two massless spinor \cite{KLEI}.
If we choose the two light-like momenta $p_{1},~p_{2}$ that build up $p$:
%---------------------------------------------------------
\bea
p_{1}^{\mu} ~+~ p_{2}^{\mu} ~=~ p^{\mu},~~  
p_{1}^{2} & = & p_{2}^{2} ~=~ 0.
\eea
%---------------------------------------------------------
we can get the complete set of massive (anti) fermion as follows.
\bea
u_{ \uparrow }( p ) & = & \left| p_{1} + \right > + 
                          t(p_{2},p_{1})\left | p_{2} - \right > 
~,~
u_{ \downarrow }( p ) =  \left | p_{1} - \right > + 
                           s(p_{2},p_{1}) \left | p_{2} + \right >, 
\\
v_{ \uparrow }( p ) & = & \left | p_{1} - \right > - 
                          s(p_{2},p_{1}) \left | p_{2} + \right >
~,~
v_{ \downarrow }( p ) =  \left| p_{1} + \right > - 
                            t(p_{2},p_{1})\left| p_{2} - \right >. 
\eea
%------------------------------------------------------
Here, we define s and t as follows:
\bea
s(p_{1},p_{2})~ & = & \frac{ \left < p_{1} + | p_{2} - \right >}{m}
               ~ = ~ - \frac{ \left < p_{2} + | p_{1} - \right >}{m}, 
\\
t(p_{1},p_{2})~ & = & \frac{ \left < p_{1} - | p_{2} + \right >}{m}
               ~ = ~ - \frac{ \left < p_{2} - | p_{1} + \right >}{m}, 
\\
| s(p_{1},p_{2})|^{2} & = & | t(p_{1},p_{2})|^{2}=1.
\eea 
The photon polarization vectors with momentum $k$ can be also written 
in terms of the massless spinor.
%
%------------------------------------------------------
\bea
\varepsilon^{\mu}_{R,L} \left( k,q \right)
& = &
\frac{\pm
      \left < q \pm| \gamma^{\mu} | k\pm  \right > 
     }{\sqrt{2} \left <k \pm|q \mp \right > 
      },
\eea
%---------------------------------------------------------
where the suffix R/L denotes the right-handed/left-handed polarization 
of the photon.
The momentum q is a reference momentum which can be taken arbitrarily.

\clearpage
%%%%%%%%%%%%%%%%%%%%%%%%%%%%%%%%%%%%%%%%%%%%%%%%%%%%%%%
\section{Appendix B}
%%%%%%%%%%%%%%%%%%%%%%%%%%%%%%%%%%%%%%%%%%%%%%%%%%%%%%
We present the analytical form of the squared
amplitudes in the CM frame for the top quark pair production from 
the two gluons ($gg$) with the spin basis for the top quark and 
the anti-top quark defined in the Sec.II. 

The squared amplitudes which correspond to the initial total angular
momentum $J = 2$ are,
%%%%%%%%%%%%%%%%%%%%%%%%%
%$g_{R} g_{L}$ to $t \bar{t}$ amplitude
%%%%%%%%%%%%%%%%%%%%%%%%%
\bea
\sum_{av}| M
\left(
        g_{R}~ g_{L} 
        \rightarrow t_{\uparrow} \bar{t}_{\uparrow} ~\mbox{and}~
t_{\downarrow} \bar{t}_{\downarrow}
\right) |^{2}
& = &
y_{gg}\left( \beta ,\theta^{\ast} \right)
\times
\beta^{2} \sin^{2}\theta^{\ast}
\\ \nonumber
&\times&
\left(
\sqrt{1-\beta^{2}}\sin\theta^{\ast}
\cos\xi 
- 
\cos\theta^{\ast} 
\sin\xi \right)^{2}  ,\\
\sum_{av}| M
\left(
g_{R}~ g_{L} 
\rightarrow t_{\uparrow} \bar{t}_{\downarrow} ~\mbox{or}~
t_{\downarrow}\bar{t}_{\uparrow}
\right) |^{2}
&=&
y_{gg}\left( \beta ,\theta^{\ast} \right)
\times
\beta^{2}\sin^{2}\theta^{\ast}
\\ \nonumber 
&\times&
\left(
\sqrt{1-\beta^{2}}\sin\theta^{\ast}
\sin\xi 
+ 
\cos\theta^{\ast}
\cos\xi\mp 1
\right)^{2}.
\eea
The ones which correspond to the initial total angular momentum $J = 0$ are,
%%%%%%%%%%%%%%%%%%%%%%%%
% $g_{R }g_{R}$ to $t \bar{t}$ amplitude 
%%%%%%%%%%%%%%%%%%%%%%%%
%
\bea
\sum_{av}| M
\left(
g_{R}~ g_{R} 
\rightarrow t_{\uparrow} \bar{t}_{\uparrow} ~\mbox{or}~
t_{\downarrow}\bar{t}_{\downarrow}
\right) |^{2}
&=&
y_{gg}\left( \beta ,\theta^{\ast} \right)
\times
\left(1-\beta^{2}\right) 
\left(1\mp \beta \cos\xi\right)^{2} ,\\
\sum_{av}| M
\left(
g_{R}~ g_{R} 
\rightarrow t_{\uparrow} \bar{t}_{\downarrow} ~\mbox{and}~
t_{\downarrow}\bar{t}_{\uparrow}
\right) |^{2}
&=&
y_{gg}\left( \beta ,\theta^{\ast} \right)
\times
\left(1-\beta^{2}\right) \beta^{2} \sin^{2} \xi.
\eea
Where $\sum_{av}$ denotes the average (summation) over 
the color indices of the initial (final) particles. 
The coefficient $y_{gg}$ is defined as follows,
\bea
y_{gg} \left( \beta ,\theta^{\ast} \right)
& \equiv &
\frac{
        \left( 4\pi\alpha_{S} \right)^{2}
        \left( 7 + 9 \beta^2 \cos^{2}\theta^{\ast} \right) 
     }{
48 
        \left( 1- \beta^{2} \cos^{2}\theta^{\ast} \right)^{2}
      }.
\eea
The calculations are somewhat complicated, but the final answers are very 
simple. Our results are consistent with the recent work \cite{GREG,PRIV}.
As mentioned previously, the squared amplitude has the same form as 
the one of $\gamma\gamma$ fusion. One difference is the form of 
the function $y_{gg}(\beta,\theta^{*})$, so the non-abelian nature of 
the $gg$ fusion appears only in this function.
 
The other squared amplitude can be obtained with use of the following 
relation.
\bea
| M
\left(  \gamma_{i}\gamma_{j} 
        \rightarrow t_{l} \bar{t}_{m}
\right) 
|^{2} & = &
| M 
\left(  \gamma_{-i}\gamma_{-j} 
        \rightarrow t_{-l} \bar{t}_{-m}
\right) |^{2},
\eea
where the suffices $i,~j$ denote the photon helicities $R$ or $L$ and
$l~(m)$ denotes the top (anti-top) quark spin state.
$(-)$ is short hand notation of the operation which change 
the $R (L,~\uparrow,~\downarrow) $ to the $L(R,~\downarrow,~\uparrow)$
state.

%
%%%%%%%%%%%%%%%%%%%%%%%%%%%%%%%%%%%%%%%%%%%%%%%%%%%%%%%
%                  References 
%%%%%%%%%%%%%%%%%%%%%%%%%%%%%%%%%%%%%%%%%%%%%%%%%%%%%%%

\clearpage
%----------------- Fig. Caption ---------------------
%%%%%%%%%%%%%%%%%%%%%%%%%%%%%%%%%%%%%%%%%%%%%%%%%%%%%%%%%%%
%%%%                  Figure                            %%%%
%%%%%%%%%%%%%%%%%%%%%%%%%%%%%%%%%%%%%%%%%%%%%%%%%%%%%%%%%%%%
%
\begin{figure}[h]
\begin{center}
\leavevmode\psfig{file=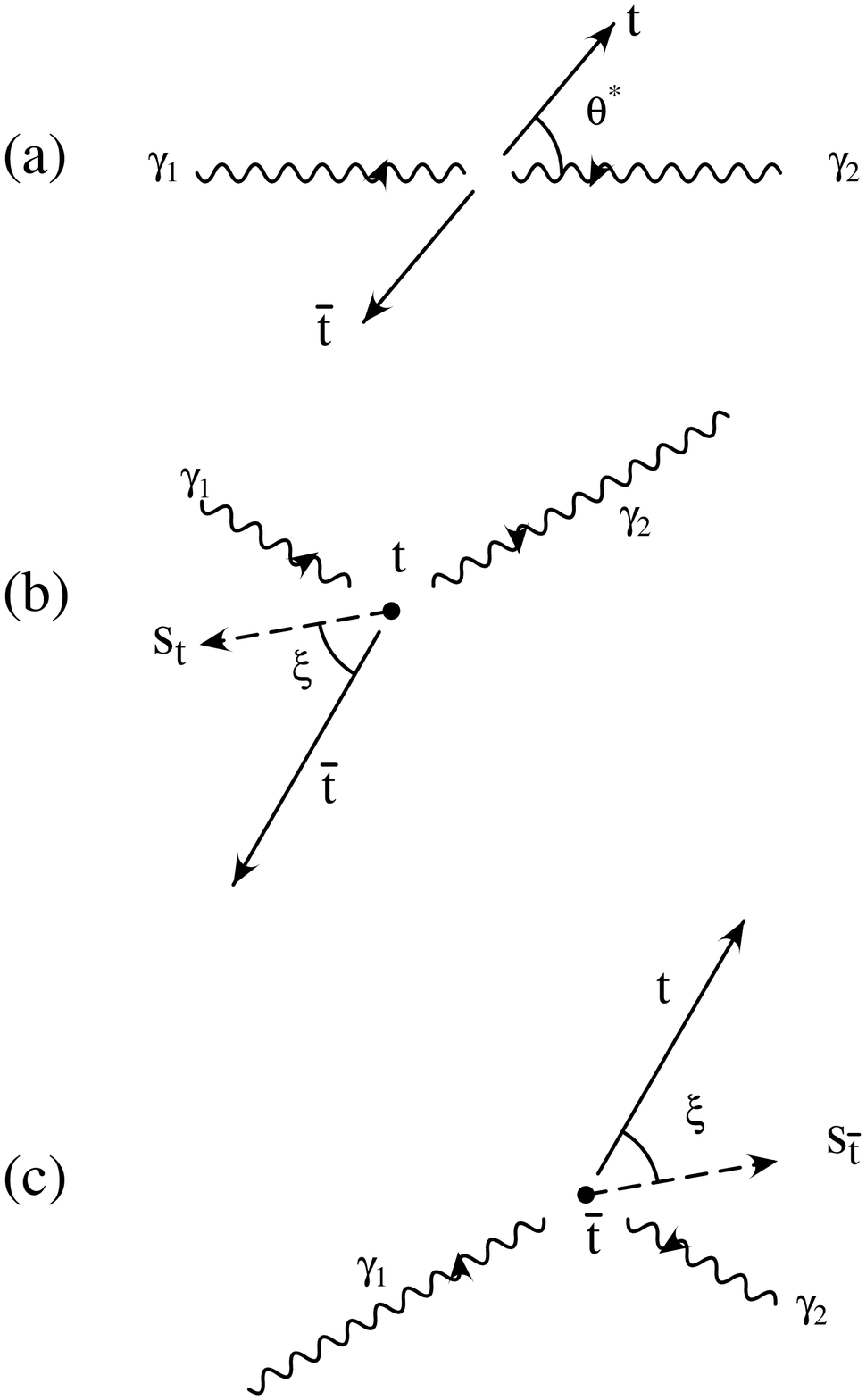,width=10cm}
\vspace{1cm}
\caption{ (a) Scattering in the CM frame.
(b) Top quark spin is defined along the direction $\vec{s}_{t}$ 
in the rest frame of the top quark, which has a relative angle $\xi$ 
to the anti-top quark momentum.  
(c) Anti-top quark spin is also defined along the direction 
$\vec{s}_{\bar{t}}$ in the rest frame of the anti-top quark.}
\end{center}
\label{fig:spin-basis}
\end{figure}
\clearpage
%%%%%%%%%%%%%%%%%%%%%%
\begin{figure}[h]
\begin{center}
\leavevmode\psfig{file=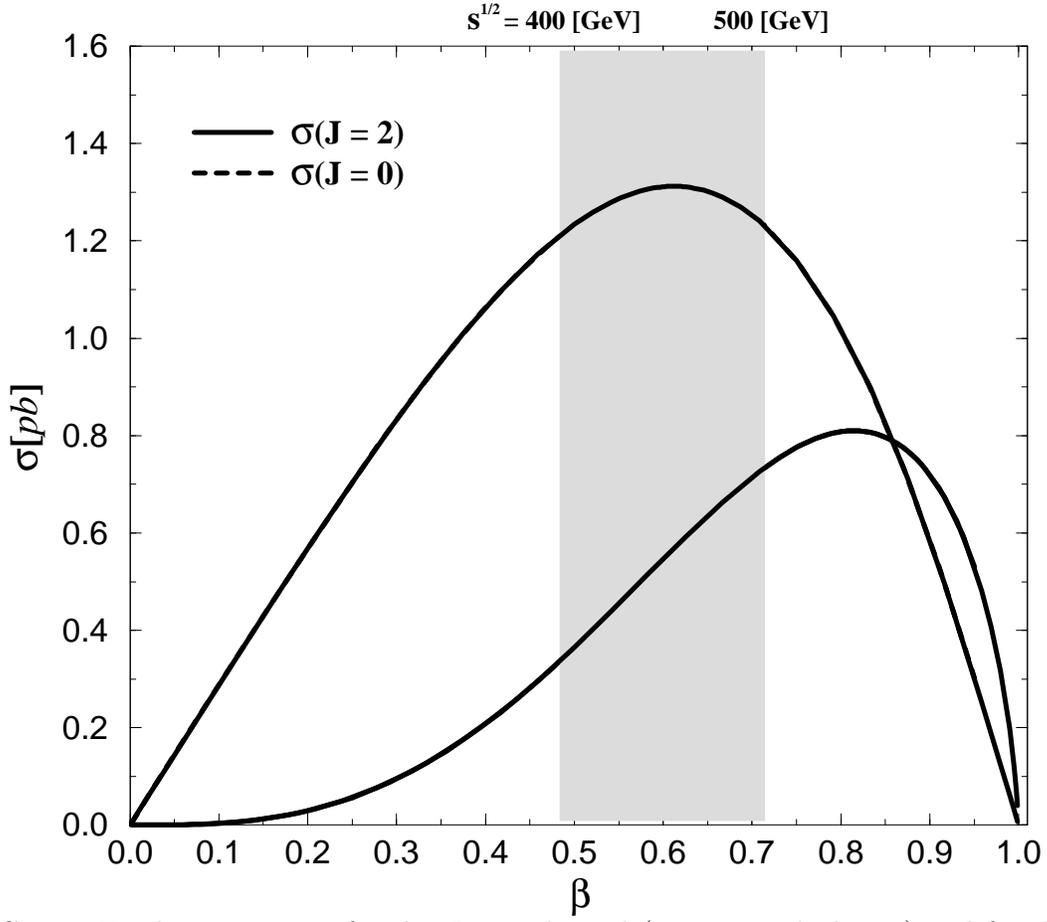,width=16cm}
\vspace*{-8cm}
\caption{ 
Total cross-section for the $J=0$ channel 
($\gamma_{R} \gamma_{R}$ initial photon) and for the 
$J=2$ channel ($\gamma_{R} \gamma_{L}$ initial photon).
The top quark spins are summed over.
}
\end{center}
\label{fig:total}
\end{figure}
\clearpage
%%%%%%%%%%%%%%%%%%%%%%
\begin{figure}[h]
\begin{center}
\leavevmode\psfig{file=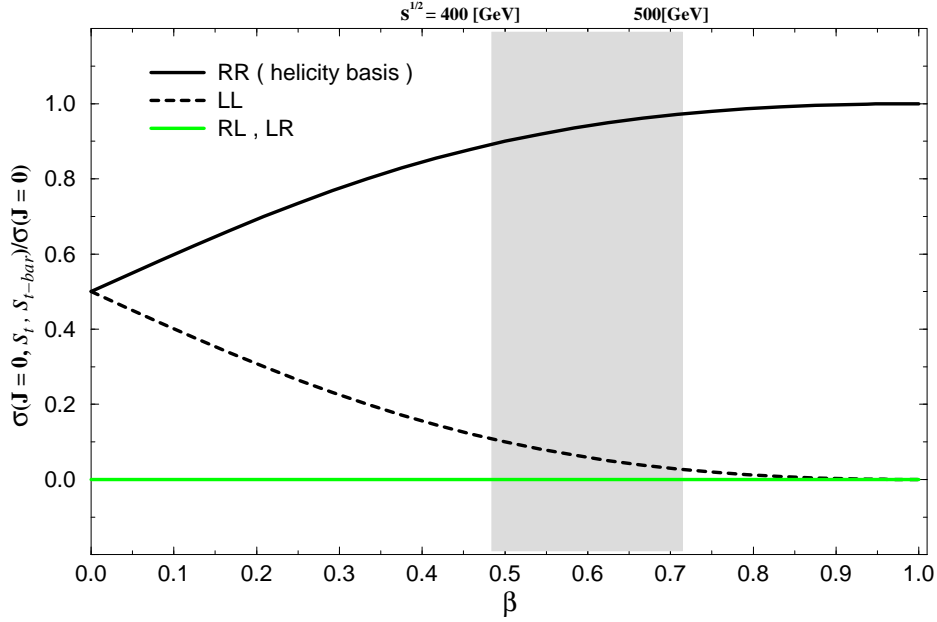,width=15cm,angle=-90}
\vspace*{-1.5cm}
\caption{
The fraction of the spin configuration (in the helicity basis)
in $J=0$ channel ($\gamma_{R} \gamma_{R}$) as a function of the speed $\beta$. 
This ratio is normalized by the total spin configuration. 
For instance upper solid line (RR) gives $\sigma(\gamma_{R}\gamma_{R} \rightarrow 
t_{R}\bar{t}_{R})/\sigma(\gamma_{R}\gamma_{R}\rightarrow total)$.
}
\end{center}
\label{fig:totRR}
\end{figure}
%\clearpage
%%%%%%%%%%%%%%%%%%%%%%
\vspace*{-1.5cm}
\begin{figure}[h]
\begin{center}
\leavevmode\psfig{file=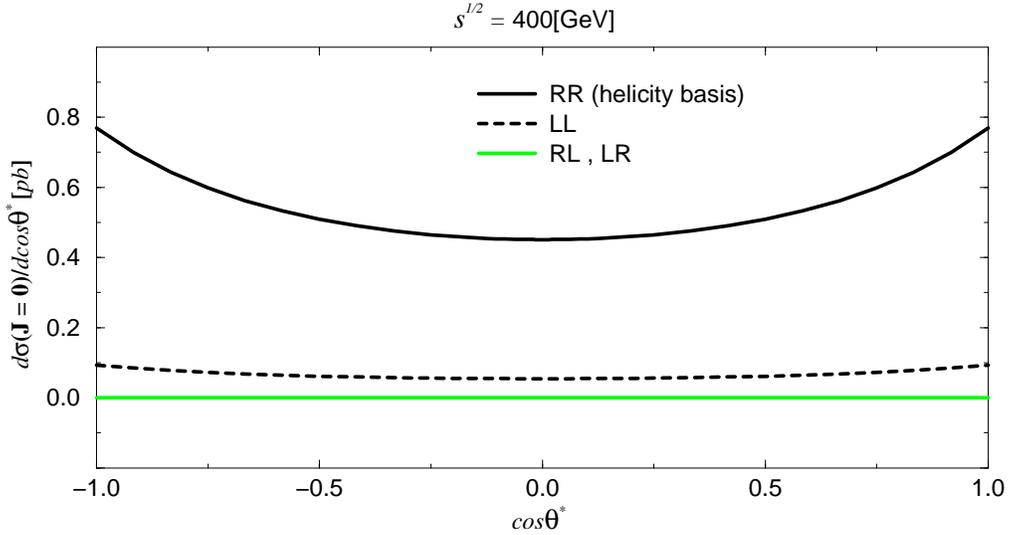,width=16cm,angle=-90}
\vspace*{-2.5cm}
\caption{ 
The differential cross-section for the process 
$\gamma_{R}\gamma_{R}\rightarrow t(s_{t}) \bar{t}(s_{\bar{t}})$ at 
$\protect\sqrt{s}=400$ GeV. 
Each lines corresponds to the the spin configuration $t(R)\bar{t}(R)$, 
$t(L)\bar{t}(L)$ and $t(R)\bar{t}(L)$.
}
\end{center}
\label{fig:difRR}
\end{figure}
\clearpage
%%%%%%%%%%%%%%%%%%%%%%
\begin{figure}[h]
\begin{center}
\leavevmode\psfig{file=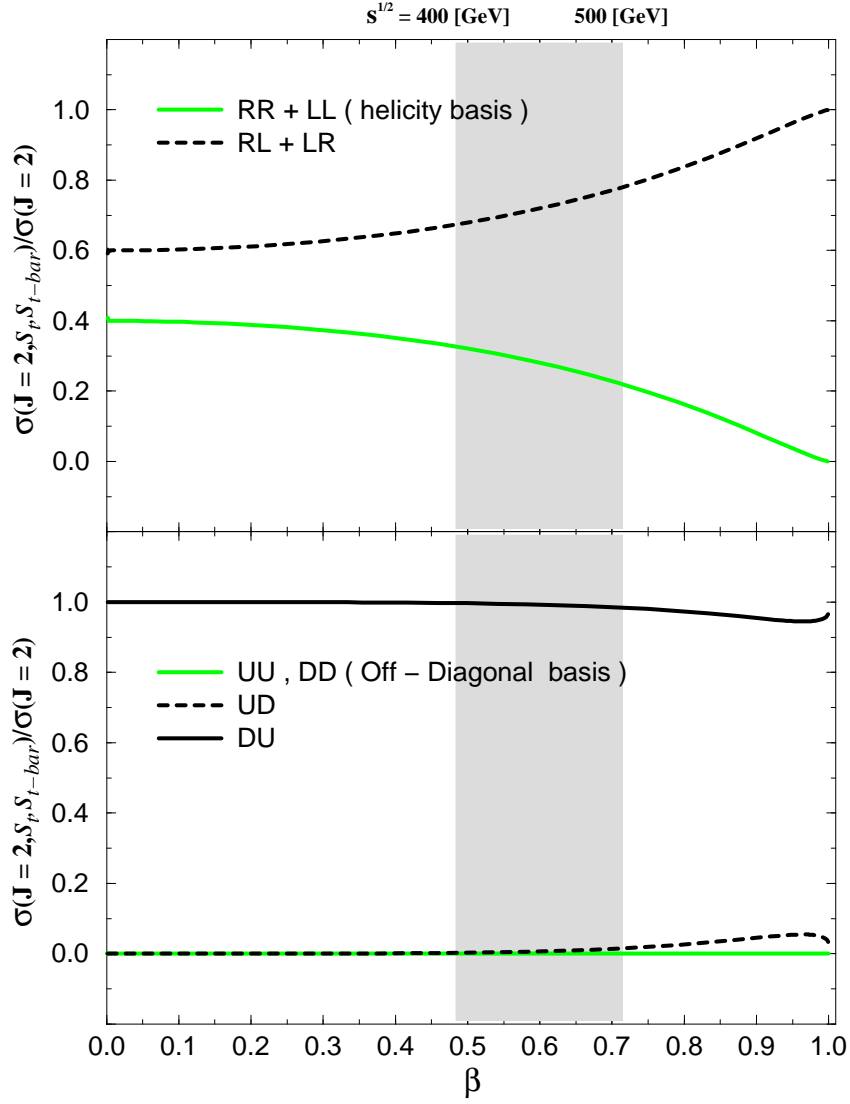,width=13cm}
%\vspace{0.5cm}
\caption{
The fraction of the spin configuration in $J=2$ channel 
($\gamma_{R} \gamma_{L}$) in the helicity and off-diagonal bases: 
RR+LL ($t_{\uparrow} \bar{t}_{\uparrow}+ t_{\downarrow}
\bar{t}_{\downarrow}$), 
RL+LR ($t_{\uparrow} \bar{t}_{\downarrow}+ t_{\downarrow} \bar{t}_{\uparrow}$)
in the helicity basis: UU+DD ($t_{\uparrow} \bar{t}_{\uparrow}+
                              t_{\downarrow} \bar{t}_{\downarrow}$), 
UD ($t_{\uparrow} \bar{t}_{\downarrow}$), 
DU ($t_{\downarrow} \bar{t}_{\uparrow}$) in the off-diagonal basis.
}
\end{center}
\label{fig:totRL}
\end{figure}
\clearpage
%%%%%%%%%%%%%%%%%%%%%%
\begin{figure}[h]
\begin{center}
\leavevmode\psfig{file=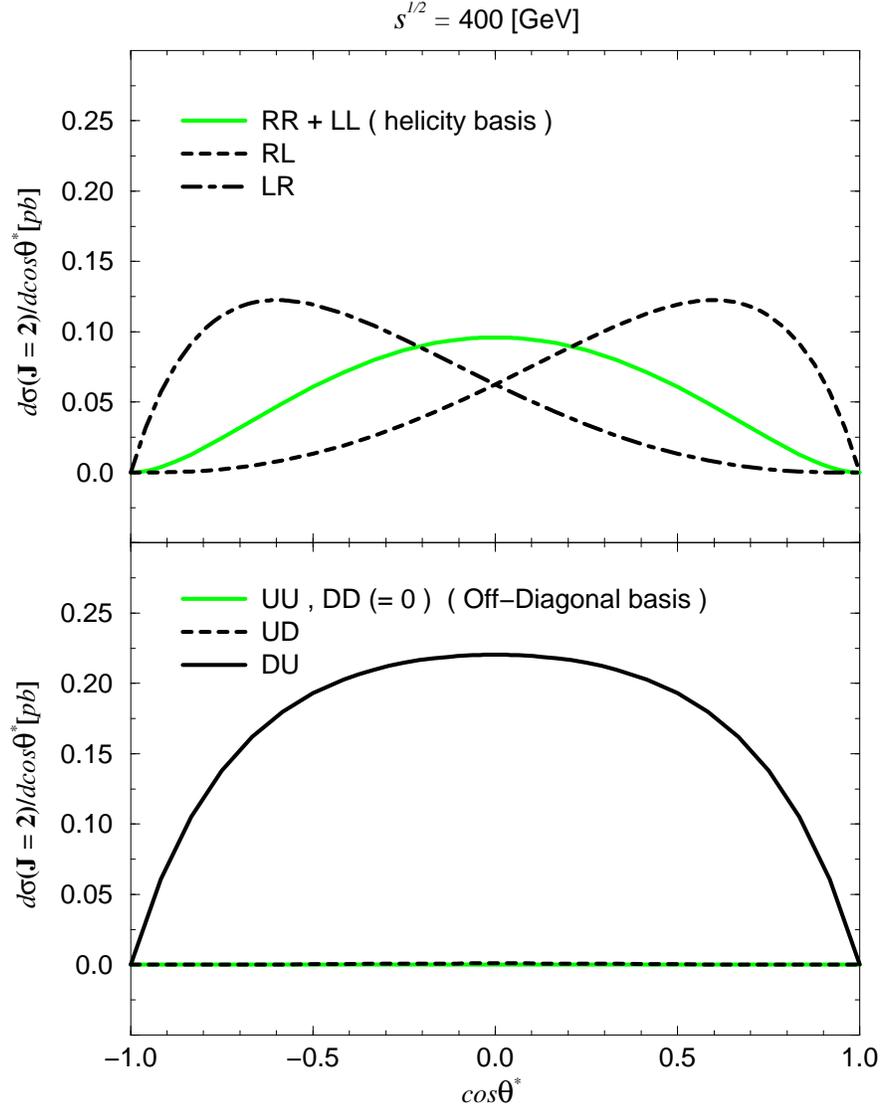,width=13cm}
%\vspace{0.5cm}
\caption{ 
The differential cross-section for the $J=2$ channel ($\gamma_{R} \gamma_{L}$)
in the helicity and off-diagonal bases: 
RR+LL ($t_{\uparrow} \bar{t}_{\uparrow} 
       + t_{\downarrow} \bar{t}_{\downarrow}$), 
RL ($t_{\uparrow} \bar{t}_{\downarrow}$), 
LR ($t_{\downarrow} \bar{t}_{\uparrow}$) in the helicity basis: 
UU+DD ($t_{\uparrow} \bar{t}_{\uparrow}+ t_{\downarrow}\bar{t}_{\downarrow}$),
UD ($t_{\uparrow} \bar{t}_{\downarrow}$), 
DU ($t_{\downarrow} \bar{t}_{\uparrow}$) in the off-diagonal basis.
}
\end{center}
\label{fig:difRL}
\end{figure}
\clearpage
%%%%%%%%%%%%%%%%%%%%%
\end{document}